\documentclass[12pt]{article}
\usepackage[cp1250]{inputenc}
\usepackage[T1]{fontenc}
\usepackage{amssymb}
\usepackage{amsmath}
\usepackage{graphicx}
\usepackage{graphics}
\usepackage{float}

\usepackage{tikz}
\usepackage{pgfmath}
\usetikzlibrary{topaths}
\usetikzlibrary{shapes.geometric}
\usetikzlibrary{calc}

\title{Estimating the Hubbard repulsion sufficient for the  onset of nearly-flat-band ferromagnetism}

\author{\L ukasz Andrzejewski and Janusz J\c{e}drzejewski \thanks{Corresponding author: J.J., jjed@ift.uni.wroc.pl}\\
Institute of Theoretical Physics, University of Wroc\l aw,\\
pl. Maksa Borna 9, 50--204 Wroc\l aw, Poland}

\begin{document}
\maketitle

\begin{abstract}
We consider nearly-flat-band Hubbard models of a ferromagnet, that is the models that are weak perturbations
of those flat-band Hubbard models whose ground state is ferromagnetic for any nonzero strength $U$ of the
Hubbard repulsion. In contrast to the flat-band case, in the nearly-flat-band case the ground state,
being paramagnetic for $U$ in a vicinity of zero, turns into a ferromagnetic one only if $U$ exceeds some
nonzero threshold value $U_{th}$.
We address the question whether $U_{th}$ of the considered models is in a physical range, therefore we
attempt at obtaining possibly good estimates of the threshold value $U_{th}$.
A rigorous method proposed by Tasaki is extended and the resulting estimates are compared with small-system,
finite-size scaling results obtained for open- and periodic-boundary conditions. Contrary to suggestions in
literature, we find the latter conditions particularly useful for our task.
\end{abstract}

\section{Introduction}
The origin of ferromagnetism in itinerant electron systems, that is of the existence of a net macroscopic
magnetic moment, without any external magnetic field, is an old problem set up already by founding fathers
of quantum mechanics.
Following almost equally old hints by Heisenberg \cite{Heisenberg-28},
it is the Fermi statistics and the Coulomb interaction between the electrons that are responsible
for this phenomenon. In their quest for ferromagnetism in itinerant electron systems,
Mielke and Tasaki (see \cite{Tasaki-93} and references quoted there) have analyzed
the paradigmatic Hubbard models that describe strongly-correlated electron systems. The general Hamiltonian
of those models reads:
\begin{eqnarray}
 H&=&\sum_{i,j,\sigma} t_{i,j}\,\,  C_{i,\sigma}^{\dagger}C_{j,\sigma}
+ \frac {1}{2}U \sum_{i,\sigma} n_{i,\sigma} n_{i,-\sigma},
\label{hubbard}
\end{eqnarray}
where the sums are over all the sites $i,j$ of the underlying lattice, and over projections
of the electron spin $\sigma$ on some axis. The operators $C_{i,\sigma}^{\dagger}$, $C_{i,\sigma}$,
stand for the creation and annihilation operator, respectively,
of an electron whose spin projection is $\sigma$, being in a state $|i\rangle$, belonging to an orthonormal basis
of the single-particle state space, that is localized at (and therefore labeled by) lattice sites $i$.
The coefficients $t_{i,j}$ -- the matrix elements of a single-particle Hamiltonian between states
$|i\rangle$ and $|j\rangle$ -- give the hopping intensities (if $i \neq j$) and on-site external
potentials (if $i=j$).
The term proportional to $U>0$ is a two-body interaction that represents a strongly screened Coulomb repulsion;
$n_{i,\sigma}= C_{i,\sigma}^{\dagger} C_{i,\sigma}$. By varying the  underlying lattices, the matrices $t_{i,j}$,
and the number of electrons, $N_e$, we obtain different systems of itinerant electrons in tight-binding approximation.
From the  physical point of view, the Hamiltonians (\ref{hubbard}) describe extremely simplified systems of
itinerant electrons. However, these models are not too simple to be relevant for the question of ferromagnetism;
the crucial property, that is the invariance with respect to rotations in the total-spin space
($SU(2)$ invariance in the total-spin space), holds true.

Apparently, the two ``ingredients'' of ferromagnetism, suggested by Heisenberg, are built in these models.
For suitable choices of the underlying lattice, the matrix elements $t_{i,j}$, and the electron number $N_e$,
Mielke and Tasaki have succeeded in proving \cite{Tasaki-93}, for the first time, that the ground state is
{\em ferromagnetic}, that is the total spin, $S_{tot}$, is proportional to the electron number $N_{e}$, for any $U>0$ .
Moreover, that ground state is a {\em saturated ferromagnet} (that is, for given $N_{e}$, $S_{tot}$ is maximal).

Their results were the first ones providing us with some fine insight into the nature of itinerant ferromagnetism,
with some sufficient conditions for ferromagnetism. Those results go far beyond the mean-field theory that leads to
the qualitative Stoner criterium \cite{Nolting-Ramakanth}, a kind of necessary condition for ferromagnetism.
In their analysis, besides the Fermi statistics and Coulomb interactions,
another factor played a crucial role -- the topology of the underlying lattice. Without going into details,
this topology should enable one to adjust the hopping intensities and external potentials in such a way that
the lowest band in the single-particle spectrum is non-dispersive, or briefly {\em flat}, and there exists a basis of
the flat-band eigensubspace, which consists of {\em localized} (i.e with bounded support) eigenstates with suitable
overlap and connectivity properties.

Now, it has to  be mentioned that the theoretical limit of flat band results in a nonphysical system that cannot
be realized in any experiment. This is  because such an electron system violates a number of physical principles.
For instance, in a range of number densities, the residual entropy is finite (see \cite{OD-1} for examples where
it was calculated exactly), and there is no Fermi surface. Moreover, the density of states of flat-band systems is
singular and there is no competition between the kinetic energy and the Coulomb interaction energy.
Therefore, as far as the physical phenomenon, the itinerant ferromagnetism, is concerned, the described
models would amount for not more than toy models, if not the subsequent work of Tasaki (\cite{Tasaki-03} and
references quoted there).

Tasaki has been able to prove that the flat-band ferromagnetism of specific Hubbard models is robust against
perturbations that broaden the flat band. At least at some cases, the perturbed flat-band models, the so called
{\em nearly-flat-band models}, have also ferromagnetic ground states but only if the value of the Hubbard repulsion
$U$ is greater than some threshold value. For sufficiently large systems and narrow broadened flat bands that
threshold value does not depend on the kind of perturbation and boundary conditions applied to the system, and is,
therefore, an intrinsic characteristic of nearly-flat-band electron systems; in the sequel denoted $U_{th}$.
This quantity appears to be the greater the stronger the perturbation is, for perturbations that are not too
strong, naturally. In contrast to flat-band models, the defined nearly-flat-band models can, in principle,
be realized in experiments.

The stability proof was a real breakthrough, despite the fact that some numerical evidence and non-rigorous arguments
for the stability of ferromagnetism in such models had been earlier obtained \cite{Kusakabe-94,Tasaki-94,Tasaki-96}.
The only comparable result, but referring to an unsaturated ferromagnetism (sometimes referred to as ferrimagnetism),
was obtained by E. Lieb \cite{Lieb-89}.
While it is hard to overestimate the importance of the above mentioned results for our understanding of ferromagnetism,
one should bear in mind that the described phenomenon is rather fragile, in the sense that it is very sensitive
to details of the underlying lattice and matrix $t_{i,j}$. In the space of the parameters of  Hamiltonian
(\ref{hubbard}), even general (not necessarily ferromagnetic) nearly-flat-band systems are of ``measure zero''.

There are many theoretical examples of nearly-flat-band systems, yet we do not know of any measurements
performed on real nearly-flat-band systems. There have been a few proposals of experimental realizations
of nearly-flat-band systems, as atomic quantum wires \cite{Arita-98}, quantum-dot super-lattices \cite{Tamura-02}
or organic polymers \cite{Suwa-03}. However, the most promising seems to be a realization  as an ultracold gas of
atoms in an optical lattice \cite{Jo-09}. Due to a very good control of system parameters in the latter case,
such a realization would open new possibilities of investigating the mechanism of nearly-flat-band ferromagnetism,
inaccessible in other experimental realizations and beyond the scope of present-day theoretical methods.

Most of the models of nearly-flat-band ferromagnets considered in literature, in particular those studied in this paper,
represent insulators or semiconductors \cite{Tasaki-93}. Some physicists might consider this feature as a drawback of
those models. However, recent investigations show that ferromagnetic insulators or semiconductors are very promising
materials for vigorously developing spintronics \cite{Jaworski-10}. Therefore, the mentioned feature constitutes another
good reason to study those nearly-flat-band electron systems that are non-metallic.

From the perspective of experimental realizations of nearly-flat-band ferromagnetic systems, it is important to know
not only the {\em flat-band conditions} on $t_{i,j}$, which guarantee that the lowest band is flat, but also to have
an estimate of the threshold value of the Hubbard repulsion, $U_{th}$, for given widths of the broadened flat band,
which are related to the strength of a perturbation.
It might happen that, for physically realizable parameters  $t_{i,j}$, the corresponding threshold value  $U_{th}$ is
beyond a range of accessible values.
While the flat-band conditions are known for numerous lattices in $D=1,2,3$ dimensions, the estimates of $U_{th}$ are
scarse. To the  best of our knowledge, there are just two papers in literature where this question was addressed.
The first one is a rough estimate by Tasaki \cite{Tasaki-95}, based on his theorem, with no attempts to optimize it.
The second one is by Ichimura et al \cite{Ichimura-98}, and amounts to calculating
the threshold value of $U$ for a few small systems and then extrapolating the results to larger systems.
The problem is (see the sections that follow) that  a small-system threshold value  of $U$, from now on denoted
$U_{th}^{s.s.}$, depends significantly on the perturbation used to broaden the underlying flat band and on boundary
conditions applied to the system.
We show that not only the values of $U_{th}^{s.s.}$, calculated in \cite{Ichimura-98}, but also their least square
extrapolation to infinite systems are much above the value of $U_{th}$.

In this paper we extend a little the method of estimating $U_{th}$ that is based on Tasaki theorem.
We calculate also $U_{th}^{s.s.}$ for small Hubbard systems with open and  periodic-boundary conditions and find,
in particular, that contrary to what is suggested in \cite{Ichimura-98} it is not the open-boundary condition but
the periodic one that is appropriate for the task of estimating $U_{th}$.

In order to minimize the computational burden, we have performed calculations for two periodic, one dimensional,
nearly-flat-band ferromagnetic Hubbard models, whose lattice consists of a small number of sublattices. The first one,
known in literature as $\Delta$-chain or saw-tooth chain (two sublattices), belongs to the Tasaki class of models and
a version of this system has been analized by Tasaki (see \cite{Tasaki-03} and references there).
The second one (three sublattices), proposed in \cite{Watanabe-96/97} as a model of quantum wires, is off the Tasaki
class; we name it the sparse $\Delta$-chain. An attempt at estimating its $U_{th}$ has been made in \cite{Ichimura-98}.
In both mentioned models, the underlying lattice is a kind of a decorated one-dimensional lattice, which is often
visualized as a quasi-one-dimensional lattice.

The plan of the paper is as follows. In Section 2 we give some details of the two methods of estimating $U_{th}$,
used in our paper. Then, in Section 3, we present our results for the $\Delta$-chain, and in Section 4 -- the results
for the sparse $\Delta$-chain. A resume of our results can be found in Section 5. Finally, in Appendix, we provide a few
tables with numerical values of some data calculated  by us, some of which are depicted in diagrams of Sections 3 and 4.

\section{The methods of estimating $U_{th}$}
\subsection{The Tasaki method}
The Tasaki method of estimating $U_{th}$ is based on his proof that a class of nearly-flat-band Hubbard systems has
the ground state, which is a saturated ferromagnet \cite{Tasaki-03}. This method enables one to calculate a value of
the Hubbard repulsion, denoted here $\overline{U}_{th}$, above which the ground state is a saturated ferromagnet,
that is  $\overline{U}_{th}$ is a rigorous upper bound for $U_{th}$ of sufficiently large lattices.
In this subsection, we consider flat-band Hubbard models with the periodic-boundary condition, whose lowest band is flat
and is separated from higher bands by a gap.
The Hamiltonians of such models we denote by $H_{fb}(0)$, where zero stands for zero width of the flat band.
It is convenient to choose the zero of the energy scale at the energy of the flat band;
as a result, the Hamiltonian  $H_{fb}(0)$ becomes positive semi-definite.

In many models, including the two studied in this paper, all the constructions, presented below, are feasible.
Let the underlying lattice $\Lambda$ consist of $n$ sublattices, denoted ${\cal{A}}$, ${\cal{B}}$, ${\cal{C}}$, \ldots,
and let each sublattice consist of $\cal{N}$ sites (i.e. $\cal{N}$ is the number of elementary cells). Let ${\cal{T}}$
be the minimal set of lattice translations $\tau$ such that the set of translated sites,$\{\tau i\}_{\tau \in {\cal{T}}}$,
coincides with the sublattice to which $i$ belongs. ${\cal{T}}$ consists of the multiples, $\tau_{0}^{m}$,
of the elementary translation, $\tau_{0}$, $m=0,\ldots, {\cal{N}} - 1$, where $\tau_{0}^{0}$ stands for the identity.

One can construct a basis of the flat-band eigensubspace that consists of $\cal{N}$ localized, overlapping
(so in general nonorthogonal) states, with supports $\{\tau \alpha \}_{\tau \in {\cal{T}}}$,
where $\alpha$ is a bounded and connected subset of the lattice; the  union of these supports coincides with the
whole lattice. We label the states of this basis by the sites of sublattice ${\cal{A}}$ and denote the creation operator
of an electron in a state belonging to the localized basis of the flat-band eigensubspace, with spin projection $\sigma$,
by $a^{\dagger}_{i,\sigma}$, $i\in {\cal{A}}$.

After that, one can  construct a basis of the orthogonal complement of the flat-band eigensubspace that consists
of $(n-1){\cal{N}}$ localized states, $\cal{N}$ states with supports $\{\tau \beta \}_{\tau \in {\cal{T}}}$,
labeled by the sites of sublattice ${\cal{B}}$, $\cal{N}$ states with supports $\{\tau \gamma \}_{\tau \in {\cal{T}}}$,
labeled by the sites of sublattice ${\cal{C}}$, \ldots, where $\beta$, $\gamma$, \ldots, are some bounded and connected
subsets of the lattice.
The creation operators of electrons in those states with spin projection $\sigma$ are denoted by
$b^{\dagger}_{i,\sigma}$, $i\in {\cal{B}}$, $c^{\dagger}_{i,\sigma}$, $i\in {\cal{C}}$,\ldots , respectively.
By means of the defined creation operators  and their adjoints, the Hamiltonian $H_{fb}(0)$ can be written in
a manifestly positive  semi-definite form:
\begin{equation}
H_{fb}(0)= \sum_{i\in {\cal{B}},\sigma}b^{\dagger}_{i,\sigma} b_{i,\sigma} +
\sum_{i\in {\cal{C}},\sigma}c^{\dagger}_{i,\sigma} c_{i,\sigma} + \ldots
+ \frac {1}{2}U \sum_{i \in \Lambda,\sigma} n_{i,\sigma} n_{i,-\sigma},
\label{fb-hubbard}
\end{equation}
after adjusting the amplitudes of the localized states defining the creation and annihilation operators.

Apparently, the two states $\prod_{i\in {\cal{A}}}a^{\dagger}_{i,\sigma}|0\rangle$, differing by $\sigma$,
belong to the ground-state subspace of the system (\ref{fb-hubbard}) for  $N_e={\cal{N}}$.
Their total spin is maximal, $S_{tot}={{\cal{N}}}/2$,
and the value of the total-spin projection is for one of them maximal, ${{\cal{N}}}/2$, and for the other -- minimal,
$-{{\cal{N}}}/2$.
By applying to the state with the maximal total-spin projection the operator $S^{-}_{tot}$, up to ${\cal{N}}-1$ times,
we obtain $2S_{tot}-1$  states with the total spin $S_{tot}={\cal{N}}/2$ and the total-spin  projection between
$S_{tot}$ and $-S_{tot}$. All the mentioned states are linearly independent, and they constitute a basis of
a $(2S_{tot} +1)$-dimensional subspace of ${\cal{N}}$-electron state space, that we name the ferromagnetic subspace
and denote ${\cal{M}}_{ferro}$.
The proof that the ground state is a saturated ferromagnet amounts to proving that ${\cal{M}}_{ferro}$ is the
ground-state subspace.
Mielke and Tasaki succeeded in constructing specific examples of Hamiltonians $H_{fb}(0)$ (that is specific
lattices and suitable matrices $t_{i,j}$ for Hamiltonian (\ref{hubbard})(see \cite{Tasaki-93} and references quoted
therein), such that the  subspace ${\cal{M}}_{ferro}$ is the ground-state subspace for any $U>0$.
In their proof, a connectivity of the underlying  lattice and some overlap properties of the supports of the flat-band
eigenstates play a key role.

There are many ways of perturbing flat-band systems, whose ground state is a saturated ferromagnet for any $U>0$,
to get a nearly-flat-band system, whose ground state is a saturated ferromagnet only for sufficiently large $U$.
However, in Tasaki method a specific perturbation is used \cite{Tasaki-03}.
The Hamiltonian of a nearly-flat-band system,$H_{fb}(s)$, obtained by perturbing in Tasaki way the considered above
flat-band system  with Hamiltonian  $H_{fb}(0)$, reads:
\begin{equation}
H_{fb}(s)= -s \sum_{i\in {\cal{A}},\sigma} a^{\dagger}_{i,\sigma} a_{c,\sigma} +
\sum_{i\in {\cal{B}},\sigma}b^{\dagger}_{i,\sigma} b_{i,\sigma} +
\sum_{i\in {\cal{C}},\sigma}c^{\dagger}_{i,\sigma} c_{i,\sigma} + \ldots
+ \frac {1}{2}U \sum_{i \in \Lambda,\sigma} n_{i,\sigma} n_{i,-\sigma},
\label{nfb-hubbard}
\end{equation}
where $s>0$ is a parameter measuring the strength of the Tasaki perturbation that broadens the flat band;
in the models studied in the sequel $s$ will be related to the width of the broadened flat band.
The advantage of the Tasaki perturbation, with respect to other ones,
is that the ferromagnetic ground-state subspace of the flat-band case ($s=0$), ${\cal{M}}_{ferro}$, is an exact
eigensubspace of $H_{fb}(s)$, for any $s$. The corresponding eigenvalue -- the ferromagnet's energy -- amounts to
$E_{ferro}= -s \varepsilon {\cal{N}}$, where $\varepsilon$ is given by the anticommutation relation,
$\{a^{\dagger}_{i,\sigma},a_{i,\sigma}\}=\varepsilon$, for $i \in {\cal{A}}$.
Remarkably, the energy $E_{ferro}$ is the lowest eigenvalue of $H_{fb}(s)$ and ${\cal{M}}_{ferro}$ is the
ground-state subspace, provided that $s$ is small enough and $U$ is large enough.
This statement has been proved by Tasaki \cite{Tasaki-03,Tasaki-95} for models that belong to the, defined by him,
class (nowadays named the Tasaki class) of Hubbard models. Below, we outline the proof and simultaneously present the
Tasaki method of estimating $U_{th}$.

Choose a bounded and connected subset, ${\cal{S}}$, of the underlying lattice that contains $m_{\alpha}$ supports
of the flat-band localized eigenstates: $\alpha$, $\tau_{0} \alpha$,$\ldots$, $\tau_{0}^{(m_{\alpha}-1)} \alpha$;
these supports are labeled  by sites in a subset ${\cal{A}}_{{\cal{S}},\alpha}$ of sublattice ${\cal{A}}$.
Moreover, we require that ${\cal{S}}$ contains all the supports $\tau \beta$, $\tau \gamma$, \ldots, that have
non-void intersection with the union of the supports $\{\tau_{0}^{m} \alpha \}_{m=0,\ldots, m_{\alpha}-1}$.
These supports are labeled by the sites in a subset ${\cal{B}}_{{\cal{S}},\beta}$ of ${\cal{B}}$,
a subset ${\cal{C}}_{{\cal{S}},\gamma}$ of ${\cal{C}}$, \ldots, respectively.

Define an operator $P_{{\cal{S}}}$ -- the so called potential, whose support is ${\cal{S}}$, such that
$H_{fb}(s) = \sum_{\tau \in {\cal{T}}} P_{\tau{\cal{S}}}$. It has the following form:
\begin{eqnarray}
\label{potential a}
P_{{\cal{S}}} = -\frac{s}{m_{\alpha}} \sum_{i \in {\cal{A}}_{{\cal{S}},\alpha},\sigma} a^{\dagger}_{i,\sigma} a_{i,\sigma}+
\sum_{i \in {\cal{B}}_{{\cal{S}},\beta},\sigma} \left[ B(m_{\alpha}) b^{\dagger}_{i,\sigma} b_{i,\sigma} +
\frac{1}{2} U B_{U}(m_{\alpha})  n_{i,\sigma} n_{i,-\sigma} \right]  \\ \nonumber
+ \sum_{i \in {\cal{C}}_{{\cal{S}},\gamma},\sigma} \left[ C(m_{\alpha}) c^{\dagger}_{i,\sigma} c_{i,\sigma} +
\frac{1}{2} U C_{U}(m_{\alpha})  n_{i,\sigma} n_{i,-\sigma}\right]  + \ldots,
\end{eqnarray}
where the model-dependent coefficients $B(m_{\alpha})$, $C(m_{\alpha})$, \ldots, are unique.
On the other hand there are numerous choices of coefficients $B_{U}(m_{\alpha})$, $C_{U}(m_{\alpha})$, \ldots .
One can fix the latter coefficients by requiring that there is a Hubbard repulsion term at each site of ${\cal{S}}$.
We adopt this convention in our calculations carried out for specific models.
The potential $P_{{\cal{S}}}$, expressed in terms of operators $C^{\dagger}_{i,\sigma}$, $C_{i,\sigma}$, reads:
\begin{eqnarray}
\label{potential C}
P_{{\cal{S}}} &=&\sum_{i,j \in {\cal{S}},\sigma } \frac{t_{i,j}}{m_{i,j}}\,\, C^{\dagger}_{i,\sigma} C_{i,\sigma}
+ \frac {1}{2} \sum_{i,\sigma} \frac{U}{m_{i}} n_{i,\sigma} n_{i,-\sigma},
\label{potential}
\end{eqnarray}
where $m_{i,j}$ and $m_{i}$, $i,j \in {\cal{S}}$, are suitable natural numbers.
We note that ${\cal{M}}_{ferro}$ is an eigensubspace of $P_{{\cal{S}}}$ to the eigenvalue $E_{{\cal{S}},ferro} = -s\varepsilon$.

Suppose that for some value of the perturbation, $s_{0}$, and the Hubbard repulsion, $U_{0}$,
the following conditions are fulfilled:\\
(i) $E_{{\cal{S}},ferro} $ is the lowest eigenvalue of $P_{{\cal{S}}}$; let ${\cal{M}}_{{\cal{S}}}$ be the corresponding
eigensubspace,\\
(ii) any state $|\Phi \rangle \in {\cal{M}}_{{\cal{S}}}$, with $N_e$ electrons, is of the form
\begin{equation}
|\Phi \rangle = \sum_{\sigma_1, \ldots,\sigma_{m_{\alpha}}}
\prod_{i \in {\cal{A}}_{{\cal{S}},\alpha}} a^{\dagger}_{i,\sigma_i} |\Phi_{\sigma_1, \ldots, \sigma_{m_{\alpha}}} \rangle,
\end{equation}
for some states $|\Phi_{\sigma_1, \ldots ,\sigma_{m_{\alpha}}} \rangle$ with $N_e - m_{\alpha}$ electrons;
this condition implies that
\begin{equation}
|\Phi \rangle =  a^{\dagger}_{i,\sigma} |\Phi_{i,\sigma} \rangle + a^{\dagger}_{i,-\sigma} |\Phi_{i,-\sigma} \rangle ,
\end{equation}
for  any site $i \in {\cal{A}}_{{\cal{S}},\alpha}$, where $|\Phi_{i,\pm \sigma} \rangle$ are some states with $N_e - 1$
electrons,\\
(iii) $n_{i,\sigma}n_{i,-\sigma} |\Phi \rangle =0$,
for every site $i \in {\cal{S}}$ and any $|\Phi \rangle \in {\cal{M}}_{{\cal{S}}}$.\\

Then, ${\cal{M}}_{ferro}$ is the ground-state subspace of $H_{fb}(s_{0})$ for the specified value $U_{0}$ of the Hubbard
repulsion. It is worth to note here that this statement holds true as well for any $U>U_{0}$, since for such values  of
$U$, the $U$-independent subspace ${\cal{M}}_{{\cal{S}}}$ continues to be the ground-state subspace of $P_{{\cal{S}}}$.
Therefore, $U_{0}$ is a rigorous upper bound for the threshold value of $U$ for any system whose lattice $\Lambda$ contains
${\cal{S}}$, in particular for a macroscopic (or infinite) system.

If a sequence of upper bounds $\overline{U}_{th}$, obtained for a sequence of potentials whose support increases indefinitely,
is convergent, then its limit coincides with $U_{th}$ of a macroscopic system.
The conditions (i), (ii) and (iii) have been proved to hold true, using  analytic arguments,  in the case of Tasaki class
of models \cite{Tasaki-03}, for a specific potential;
this was achieved in an asymptotic region of parameters (where, in particular, $U/s$ is sufficiently large).

We note that it is enough to investigate the potential $P_{{\cal{S}}}$ in the Fock space corresponding to the system whose
support is ${\cal{S}}$, and then $P_{{\cal{S}}}$ can be represented by a finite matrix.
If the support ${\cal{S}}$ is small enough,
then a numerical exact diagonalization procedure is feasible, and the above conditions can be verified numerically.
By carrying out such a procedure for a  given value of perturbation, $s_{0}$, and a decreasing sequence of values of $U_{0}$,
an optimal estimate of $\overline{U}_{th}$ can be determined.
Eventually, one ends up with a computer assisted proof of ferromagnetism for given $s_{0}$ and any $U>\overline{U}_{th}$.

We follow the route, outlined above, i.e. calculate possibly good estimate of $\overline{U}_{th}$ for a sequence of potentials
$P_{{\cal{S}}}$ with increasing supports, first in Section 3, where a model from the Tasaki class ($\Delta$-chain) is studied,
and then in Section 4, where a model off the Tasaki class (sparse $\Delta$-chain) is considered.
In particular, in the latter case the carried out calculations amount to a computer assisted proof that the ground state of
a nearly-flat-band sparse $\Delta$-chain is a saturated ferromagnet.

Actually, the matrices that appear in our calculations are huge and sparse; their size reaches the order $10^5$. To diagonalize
such matrices we used the C library {\em Primme} of Andreas Stathopoulos and James R. McCombs  \cite{Primme},
\cite{Diagonalization 1},\cite{Diagonalization 2}.

\subsection{Small-system estimates of  $U_{th}$}

A straightforward way of estimating $U_{th}$ is to calculate it for systems on small lattices, whose number of sites,  $N_s$
constitute an increasing sequence, terminating usually at the highest level for which a numerical exact-diagonalization
procedure of the underlying Hamiltonian is feasible for our computer facilities.
We specify a small piece of the underlying lattice, the parameters of Hamiltonian (\ref{hubbard}), a boundary condition
and the electron number $N_{e}$, and carry out an exact numerical  diagonalization. The latter procedure provides
us, in particular, with the lowest eigenvalue and a basis of its eigensubspace.
In order to qualify the ground-state subspace of our small system as a saturated ferromagnet, we verify if \\
(a) the number of elements of this basis is $N_e + 1$,\\
(b) all the elements of this basis are eigenvectors of the square of the total-spin operator to the same eigenvalue
$N_e/2[(N_e/2) + 1]$  (in units where $\hbar =1$), that is the total spin of the ground-state subspace is $S_{tot}=N_e/2$.\\
Repeating the above verification procedure for a decreasing sequence of values of $U$, we determine a value of $U$,
denoted $U_{th}^{s.s.}$, that separates the values of $U$ for which the system is a saturated ferromagnet from those
for which it is not. For a given flat-band model, $U_{th}^{s.s.}$ depends on $N_s$, a kind of perturbation and boundary
conditions used.
Finally, having a sequence of the values of $U_{th}^{s.s.}$, obtained for systems of different size, we try to guess how
$U_{th}^{s.s.}$ scales with $N_s$, in order to  predict the corresponding value, $U_{th}$,
for a macroscopic system.

We end up this subsection with  a comment on the choice of the electron number $N_{e}$ in small-system calculations.
Most of the proofs of flat-band ferromagnetism and the Tasaki proof of nearly-flat-band ferromagnetism can be carried out
only for particular $N_{e}$, such that exactly half of a band in the single-particle spectrum can be filled.
For periodic boundary conditions such a $N_{e}$ amounts to $N_{s}/n$.
However, for open boundary condition this definition looses its meaning, since the bands are not well defined and, moreover,
$N_{s}$ might not be divisible by $n$. In the latter case, we choose one of the two integers that are the closest to $N_{s}/n$.
As will be demonstrated in Sections 3 and 4, the results of calculations depend significantly on $N_s$, whether it is divisible
by $n$, and on the chosen integer approximation to half-integer $N_{s}/n$.

\begin{figure}[H]
\begin{center}
\input{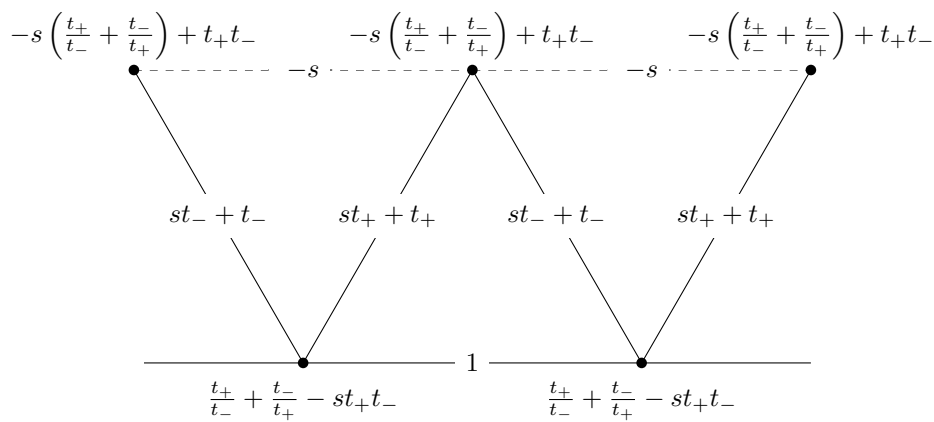}
\caption{\label{Delta chain}
The nearly-flat-band $\Delta$-chain. The expressions across the bonds of quasi 1D lattice are hopping intensities, those by the
sites are external potentials adjusted to fulfill the flat-band condition when $s=0$.}
\end{center}
\end{figure}

\section{The case of $\Delta$-chain}

The first system for which we consider the problem of estimating the threshold value $U_{th}$ of a macroscopic system
is the $\Delta$-chain. We shall argue that an extrapolation of small-system estimates obtained for periodic-boundary
condition is a reasonable approximation to $U_{th}$ of a macroscopic system.
The $\Delta$-chain is a periodic Hubbard system whose quasi 1D lattice and the parameters
of  the one-body part of Hamiltonian (\ref{hubbard}) are depicted in Fig.~\ref{Delta chain}, for the system perturbed
in Tasaki way.

\begin{figure}[H]
\begin{center}
\input{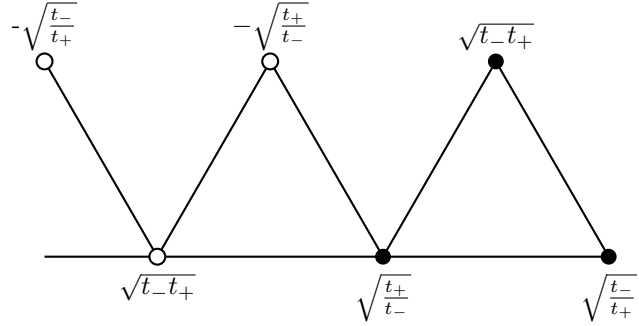}
\caption{\label{Delta chain eigenstates}
The localized flat-band eigensubspace basis -- a state $a^{\dagger}_{i,\sigma}|0\rangle$, $i \in {\cal{A}}$,
and the localized basis of the orthogonal complement of the flat-band eigensubspace -- a state
$b^{\dagger}_{i,\sigma}|0\rangle$, $i \in {\cal{B}}$.
Open circles -- the support $\alpha$ -- "V-valley",
solid circles -- the support $\beta$. The expressions by the sites are the components of those states.}
\end{center}
\end{figure}

\begin{figure}[H]
\begin{center}
\input{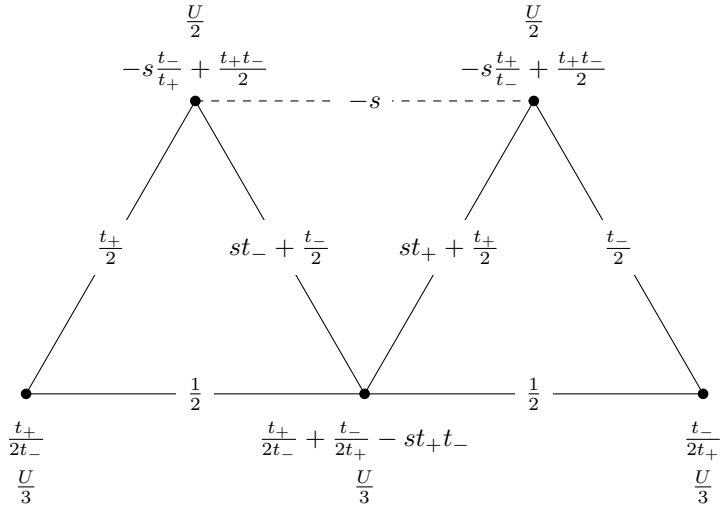}
\caption{\label{one-V-valley potential}
A potential whose support consists of 5 sites --one-V-valley potential.
The expressions across the bonds are the hopping intensities, those by the sites and independent of $U$ are external potentials,
and those proportional to $U$ are the coefficients in the Hubbard repulsive interaction .}
\end{center}
\end{figure}

The underlying lattice consists of two sublattices: one formed by
the sites in the bases of the triangles made by continuous bonds,
denoted ${\cal{A}}$, and the other -- by the tops of these
triangles, denoted ${\cal{B}}$. A bond between two sites is
represented by a continuous line, if the corresponding hopping
intensity is nonzero for any $s>0$, and by a dashed line, if it
vanishes in the unperturbed system. The hopping intensities between
neighboring sites of an elementary cell and between sites of two
such cells are independent parameters. We choose one of them as the
energy unit. In the unperturbed case ($s=0$), the hopping
intensities are $t_{+}$, $t_{-},1$, and the on-site external
potentials are adjusted to fulfill the flat-band condition: for $i
\in {\cal{A}}$, $t_{i,i}=\frac{t_+}{t_-}+ \frac{t_-}{t_+}$; for $i
\in {\cal{B}}$, $t_{i,i} = t_{+}  t_{-}$, and to set the flat-band
energy to zero.  The flat-band is the lowest one, if $t_{+}t_{-}>0$.
The width of the broadened flat band is not greater than $4s$ and
the gap above it is not smaller than
$(1+s)\left(\frac{t_+}{t_-}+\frac{t_-}{t_+}+t_+t_--2\right)>0$.

The flat-band eigensubspace of the defined system is generated by linearly independent (nonorthogonal in general)
localized states $a^{\dagger}_{i,\sigma}|0\rangle$ -- "V-valley", for $i \in {\cal{A}}$. The support of such a state,
$\alpha$, and the corresponding components are shown in Fig.~\ref{Delta chain eigenstates}.
A localized basis of the orthogonal complement of the flat-band eigensubspace consists of states denoted
$b^{\dagger}_{i,\sigma}|0\rangle$, $i \in {\cal{B}}$. The support of such a state, $\beta$, and the corresponding components
are also shown in Fig.~\ref{Delta chain eigenstates}.

\begin{figure}[H]
\begin{center}
\input{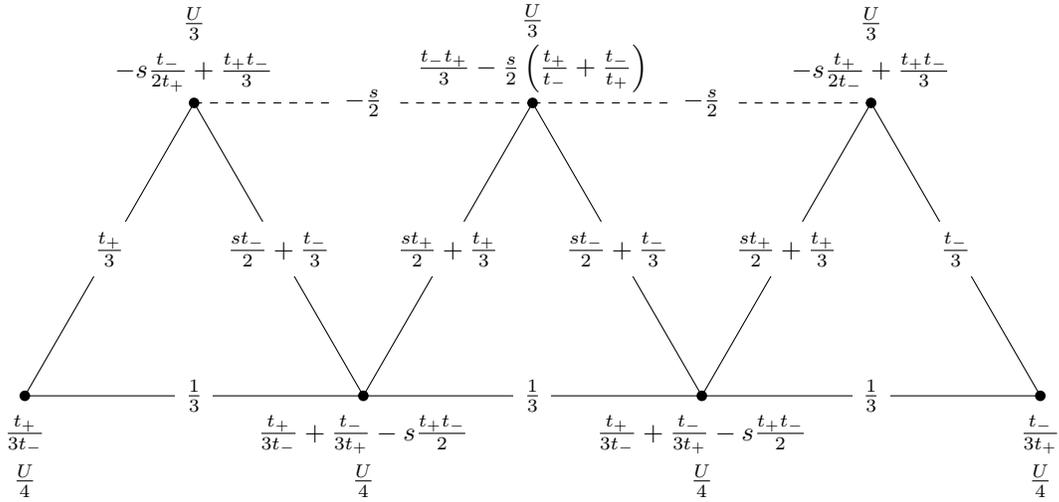}
\caption{\label{two-V-valley potential}
A potential whose support consists of 7 sites -- two-V-valley potential.
The expressions across the bonds are the hopping intensities, those by the sites and independent of $U$ are external potentials,
and those proportional to $U$ give the coefficients in the Hubbard repulsive interaction.}
\end{center}
\end{figure}

To apply the Tasaki method, we define the potentials $P_{{\cal{S}}}$:
\begin{eqnarray}
\label{delta potentials}
P_{{\cal{S}}} =  \sum_{i \in {\cal{A}}_{{\cal{S}},\alpha},\sigma}
\left[- \frac{s}{m_{\alpha}} a^{\dagger}_{i,\sigma} a_{i,\sigma} + \frac{U}{2(m_{\alpha} + 1)}   n_{i,\sigma} n_{i,-\sigma} \right]\\
\nonumber
+  \sum_{i \in {\cal{B}}_{{\cal{S}},\beta},\sigma} \left[ \frac{1}{m_{\alpha}+1} b^{\dagger}_{i,\sigma} b_{i,\sigma} +
\frac{U}{2(m_{\alpha} + 2)}   n_{i,\sigma} n_{i,-\sigma} \right].
\end{eqnarray}
Specifically, the potentials (\ref{delta potentials}) containing one-V-valley and two-V-valleys,
are given, in the form (\ref{potential C}), in Figs.~\ref{one-V-valley potential},\ref{two-V-valley potential}.

\begin{figure}[H]
\begin{center}
    \includegraphics[scale=0.4]{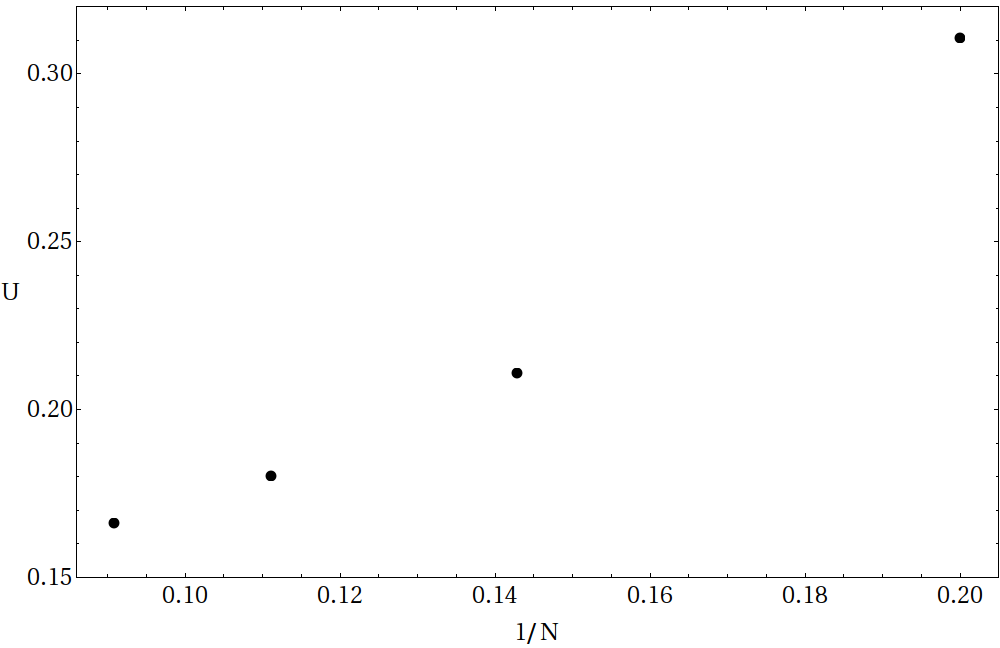}
 %   \vspace{-80pt}
    \caption{\label{thresholds from potentials}
    The case of a nearly-flat-band $\Delta$-chain, with $t_{+}=1.0$ and $t_{-}=1.5$, perturbed in Tasaki way, with $s=0.01$.
    $\overline{U}_{th}$ versus the inverse number of sites, $1/N_{{\cal{S}}}$, in the supports ${\cal{S}}$
    of the potentials $P_{{\cal{S}}}$, defined in  (\ref{delta potentials}), with $N_{{\cal{S}}}=5,7,9,11$.}
\end{center}
\end{figure}

Choosing a specific $\Delta$-chain and calculating
$\overline{U}_{th}$ for the defined potentials, whose supports
consist of $N_{{\cal{S}}}=5,7,9,11$ sites, gives us a decreasing
with $N_{{\cal{S}}}$ sequence, which might saturate at a nonzero
value, Fig.~\ref{thresholds from potentials}. The smallest
$\overline{U}_{th}$ is our best estimate of $U_{th}$ obtained by
Tasaki method.

In Fig.~\ref{potential, p.b.c., o.b.c.} we give the values of $U_{th}^{s.s.}$ obtained for small systems with open and
periodic-boundary conditions;
for comparison we include also in this figure the values of $\overline{U}_{th}$ obtained by Tasaki method.
Clearly, the values of $U_{th}^{s.s.}$ obtained for open-boundary condition are much above the rigorous upper bounds provided
by Tasaki method. The values of $U_{th}^{s.s.}$ obtained for periodic-boundary condition seem to oscillate around a constant
and are a little below the values of $\overline{U}_{th}$.

The behaviour of the latter two kinds of estimates of $U_{th}$ is more clearly seen in Fig.~\ref{potential, p.b.c.}.
It is tempting to suggest that the best (say, in the least-square fit sense) constant approximating the values of
$U_{th}$ obtained for periodic-boundary condition is a good approximation to $U_{th}$, lower than the lowest value of
$\overline{U}_{th}$, for the considered sizes of small systems and supports of the potentials.

\begin{figure}[H]
\begin{center}
    \includegraphics[scale=0.25]{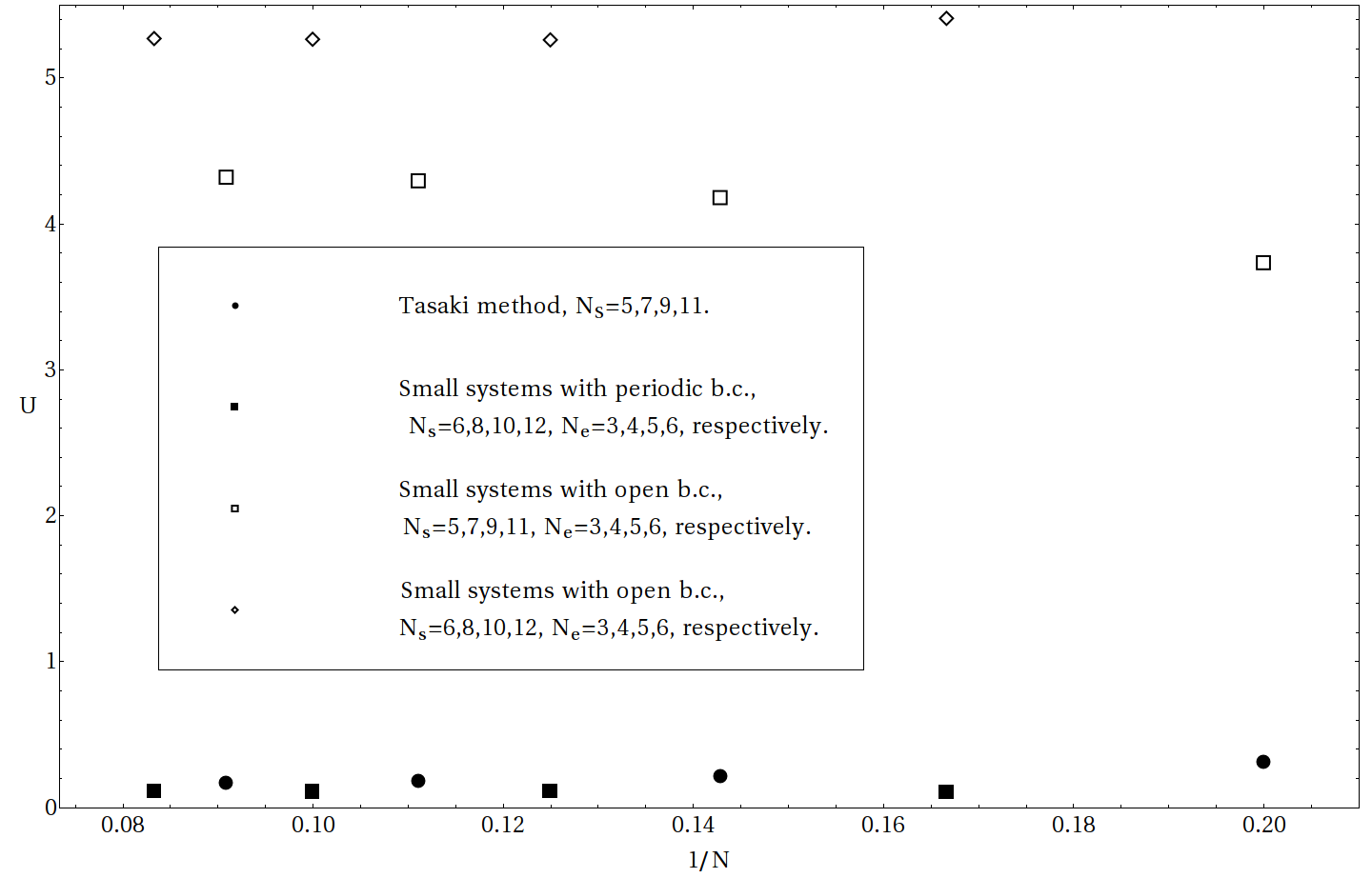}
%    \vspace{-80pt}
    \caption{
    \label{potential, p.b.c., o.b.c.}
    The case of a nearly-flat-band $\Delta$-chain, with  $t_{+}=1.0$, $t_{-}=1.5$, perturbed in Tasaki way, with $s=0.01$.
    The threshold value of $U$ versus the inverse number of sites $1/N$, i.e.
    $U_{th}^{s.s.}$ versus  $1/N_{s}$ -- for small-system data, and $\overline{U}_{th}$  versus $1/N_{{\cal{S}}}$ --
    obtained by Tasaki method with the potentials defined in (\ref{delta potentials}).
    Numerical values of the data for periodic boundary condition (periodic b.c.) are given in
    Tab.\ref{small delta-chain, Tasaki pert.,p.b.c.}.
    }
\end{center}
\end{figure}

\begin{figure}[H]
\begin{center}
    \includegraphics[scale=0.25]{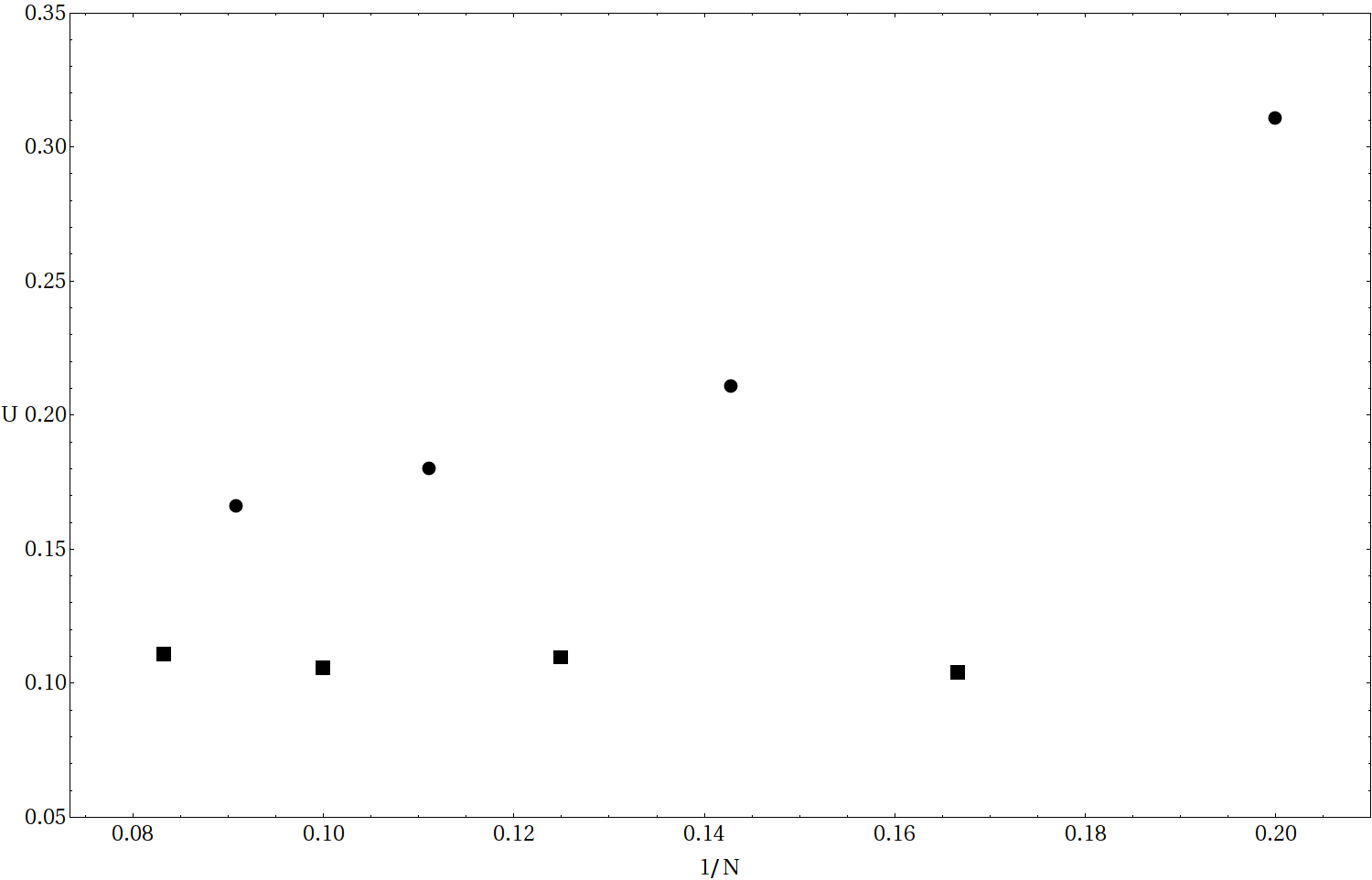}
%\vspace{-100pt}
\caption{\label{potential, p.b.c.}
    A part of data presented in Fig.(\ref{potential, p.b.c., o.b.c.}), displayed with a higher resolution.
    The case of a nearly-flat-band $\Delta$-chain, with  $t_{+}=1.0$, $t_{-}=1.5$, perturbed in Tasaki way,
    with $s=0.01$.
    The threshold value of $U$ versus the inverse number of sites $1/N$: solid squares --
    $U_{th}^{s.s.}$ versus  $1/N_{s}$ for small-systems with periodic boundary condition,
    bullets -- $\overline{U}_{th}$  versus $1/N_{{\cal{S}}}$ obtained by Tasaki method
    with the potentials defined in (\ref{delta potentials}).
    Numerical values of the data for periodic boundary condition (periodic b.c.) are given in
    Tab.\ref{small delta-chain, Tasaki pert.,p.b.c.}.
    }
\end{center}
\end{figure}

\section{The case of sparse $\Delta$-chain}

\begin{figure}[H]
\begin{center}
\input{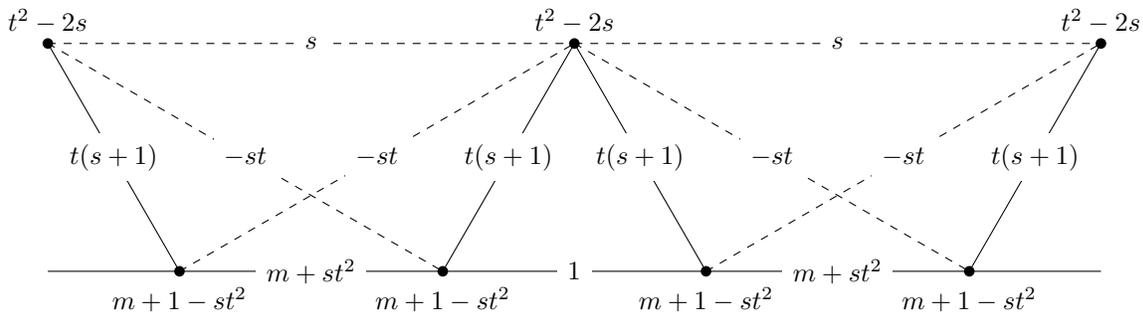}
\caption{\label{sparse Delta chain}
The nearly-flat-band sparse $\Delta$-chain, for Tasaki perturbation.
The expressions across the bonds of quasi 1D lattice are hopping intensities, those by the sites are external potentials
adjusted to fulfill the flat-band condition, when $s=0$. }
\end{center}
\end{figure}

The second system, we consider, is a modification of the $\Delta$-chain, which we name the sparse $\Delta$-chain.
This is a periodic Hubbard system whose quasi 1D lattice and the parameters of  the one-body part of Hamiltonian (\ref{hubbard}),
with Tasaki perturbation, are depicted in Fig.~\ref{sparse Delta chain}.

An attempt to estimate $U_{th}$ of a  macroscopic nearly-flat-band sparse $\Delta$-chain was made in \cite{Ichimura-98}.
The underlying lattice consists of three sublattices: the sublattice ${\cal{A}}$ that consists of the left sites of the bases
of the triangles made by continuous bonds, the sublattice ${\cal{B}}$ that consists of the the tops of those triangles,
and the sublattice ${\cal{C}}$ that consists of the right sites of the bases of those triangles.
A bond between two sites is represented by a continuous line, if the corresponding hopping intensity is nonzero for any $s>0$,
and by a dashed line, if it vanishes in the unperturbed system. There are three independent hopping intensities.
The hopping intensities between neighboring sites of the sublattice ${\cal{C}}$ and the sublattices ${\cal{A}}$ and ${\cal{B}}$
are the same, $t$, in an unperturbed system. Then, the hopping intensity between two sites belonging to the base of a triangle
made by continuous bonds is chosen as the energy unit.

Finally, the hopping intensity between two neighboring sites in the bases of neighboring triangles is $m$, in an unperturbed system.
The hoping intensities, $t,m$, are independent parameters. On the other hand, the on-site external potentials are adjusted to fulfill
the flat-band condition and to set the flat-band energy to zero: the potential at the sites of the sublattice ${\cal{C}}$ is $t^2$
and that at the sites of the sublattices ${\cal{A}}$ and ${\cal{B}}$ is $m+1$ (see Fig.~\ref{sparse Delta chain}).
We shall see below that the flat-band is the lowest one, if $m>0$ ($t$ can be of any sign).
In order to make our model the counterpart of that in \cite{Ichimura-98}, we have to set an extra relation between $t,m$:
$t=\sqrt{m}$ for $m>0$. The calculations have been carried out for $m=0.8$ (the value chosen by Ichimura et al \cite{Ichimura-98}),
and $s=0.070$. For this particular sparse-$\Delta$-chain, the width of the broadened flat band (Tasaki perturbation) is $0.136$,
which is $2s$ approximately, and the gap above the broadened flat band is $0.342$.

The support $\alpha$ --  of a localized  element of  the flat-band
eigensubspace basis, $a^{\dagger}_{i,\sigma}|0\rangle$, $i \in
{\cal{A}}$ and its components are shown in Fig.~\ref{sparse Delta
chain eigenstates}. Such a support can be depicted as a valley
between two neighboring triangles made by continuous bonds -- a
"W-valley". The supports, $\beta$ and $\gamma$, and components of
elements of a localized basis of the orthogonal complement of the
flat-band eigensubspace, $b^{\dagger}_{i,\sigma}|0\rangle$, $i \in
{\cal{B}}$, and $c^{\dagger}_{i,\sigma}|0\rangle$, $i \in
{\cal{C}}$, are also shown in Fig.~\ref{sparse Delta chain
eigenstates}.

\begin{center}
\begin{figure}[H]

\input{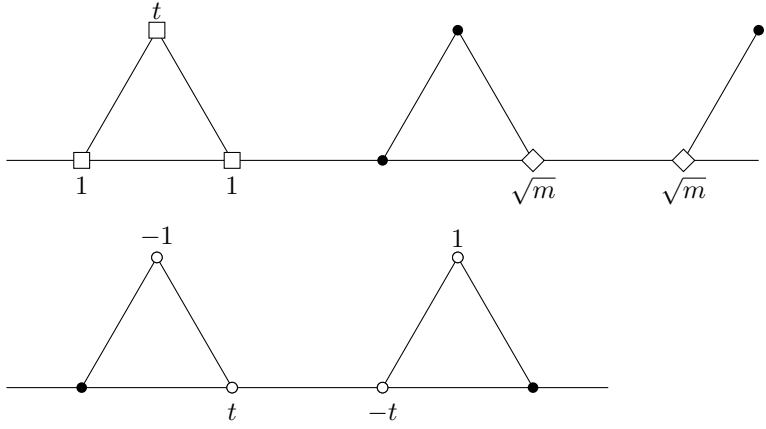}
\caption{\label{sparse Delta chain eigenstates}
Localized flat-band basis -- the states $a^{\dagger}_{i,\sigma}|0\rangle$, $i \in {\cal{A}}$, and localized basis of the
orthogonal complement of flat-band eigensubspace -- the states $b^{\dagger}_{i,\sigma}|0\rangle$, $i \in {\cal{B}}$,
and $c^{\dagger}_{i,\sigma}|0\rangle$, $i \in {\cal{C}}$.
Open circles -- the support of a localized flat-band eigenstate ("W-valley"),
open squares -- the support of  $b^{\dagger}_{i,\sigma}|0\rangle$, $i \in {\cal{B}}$,
open rhombs -- the support of  $c^{\dagger}_{i,\sigma}|0\rangle$, $i \in {\cal{C}}$.
The expressions by the sites stand for the components of those states.}
\end{figure}
\end{center}

By means of the defined above fermion creation operators and the
corresponding annihilation operators, the Hamiltonian of the
flat-band sparse $\Delta$-chain, can be written in the positive
semi-definite form (\ref{fb-hubbard}). This proves that $m>0$ is a
sufficient condition for the flat band to be the lowest one.

\begin{figure}[H]
\begin{center}
\input{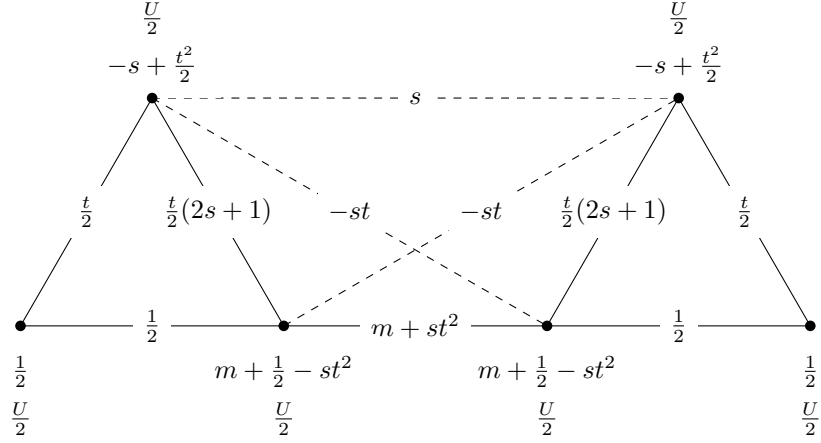}
\caption{\label{one-W-valley potential}
A potential whose support consists of 6 sites --one-W-valley potential.
The expressions across the bonds are the hopping intensities, those by the sites,
and independent of $U$, are external potentials, and those proportional to $U$ are the
coefficients in the Hubbard repulsive interaction.}
\end{center}
\end{figure}

\begin{figure}[H]
\begin{center}
\input{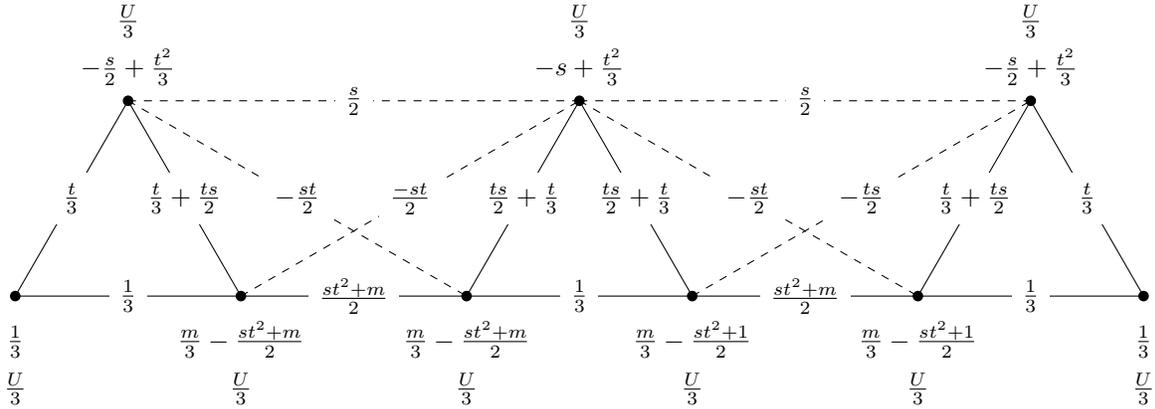}
\caption{\label{two-W-valley potential}
A potential whose support consists of 9 sites -- two-W-valley potential.
The expressions across the bonds are the hopping intensities, those by the sites,
and independent of $U$, are external potentials, and those proportional to $U$ are the
coefficients in the Hubbard repulsive interaction.}
\end{center}
\end{figure}

The potentials $P_{{\cal{S}}}$ that we use when applying Tasaki method are defined as follows:
\begin{eqnarray}
\label{sparse delta potentials}
P_{{\cal{S}}} = - \frac{s}{m_{\alpha}} \sum_{i \in {\cal{A}}_{{\cal{S}},\alpha},\sigma} a^{\dagger}_{i,\sigma} a_{i,\sigma}
 + \frac{1}{m_{\alpha}} \sum_{i \in {\cal{B}}_{{\cal{S}},\beta},\sigma}  b^{\dagger}_{i,\sigma} b_{i,\sigma}
+ \frac{1}{m_{\alpha} + 1} \sum_{i \in {\cal{C}}_{{\cal{S}},\gamma},\sigma}  c^{\dagger}_{i,\sigma} c_{i,\sigma}\\
\nonumber
+ \frac{U}{m_{\alpha} + 1} \sum_{i \in \Lambda, \sigma}  n_{i,\sigma} n_{i,-\sigma} .
\end{eqnarray}
In the form (\ref{potential C}) they are shown in Figs.~\ref{one-W-valley potential},\ref{two-W-valley potential},
for $N_{{\cal{S}}}=6,9$. One can check easily that if the conditions (i), (ii) and (iii) of the Tasaki method are satisfied
for these potentials, then the Tasaki proof of saturated ferromagnetism in the ground-state \cite{Tasaki-03}
can be carried out.

Up to now, we have dealt with only one way of perturbing a flat-band
system to obtain a nearly-flat-band one, i.e. the Tasaki
perturbation defined in (\ref{nfb-hubbard}). An alternative way
consists in modifying one or more arbitrary chosen hoping
intensities and/or external potentials, while keeping the remaining
ones intact, so that the flat-band condition is no longer valid.
This is the way adopted by Ichimura et  al in \cite{Ichimura-98} to
perturb the above defined flat-band sparse $\Delta$-chain.

Apparently, different perturbations of the same flat-band system
result in different nearly-flat-band systems. In order to compare
the threshold values of $U$ calculated for such a different systems
we have to make the systems itself comparable. We have chosen to
consider different kinds of perturbations as equivalent, or having
the same strength, if the widths of the broadened flat bands in two
nearly-flat-band systems, obtained from a given flat-band one, are
the same. To motivate our choice, we note that according to good
qualitative arguments \cite{Tasaki-96} the way of perturbing a
flat-band system is not relevant for the existence of the phenomenon
of nearly-flat-band ferromagnetism; whatever the perturbation, there
is $U_{th}>0$, which increases with the strength of perturbation.

Moreover, we expect that for sufficiently large electron systems,
with sufficiently narrow nearly-flat-bands, the threshold value of
$U$ is an increasing function of the width of the nearly-flat-band
only. In other words, $U_{th}$ is an intrinsic property of
nearly-flat-band electron systems, independent of boundary
conditions or a kind of weak perturbation applied to the underlying
flat-band system; the only relevant physical parameter,
characterizing a sufficiently narrow nearly-flat-band, is its width.

In \cite{Ichimura-98}, the flat-band sparse $\Delta$-chain is perturbed  by replacing the hoping intensity $t=\sqrt{m}$ by
$t=\sqrt{m} + \delta$, for some $\delta \neq 0$ (the Ichimura et al perturbation). Now, we are ready to compare estimates of
$U_{th}$ for two nearly-flat-band sparse $\Delta$-chains, one obtained  by perturbing the flat-band sparse $\Delta$-chain
in  Tasaki way and the other -- in Ichimura et al way; by adjusting the parameter $s$ of Tasaki perturbation we can make equal
the widths of the broadened flat bands in the both cases. Specifically, for $m=0.8$ and $t=1$  the width of the broadened flat
band (Ichimura et al perturbation) is $0.136$, and with Tasaki perturbation the same width is attained for $s=0.070$.
The gap above the broadened flat band is $0.210$.

In Fig.~\ref{potential, ichimura p.b.c. and o.b.c., tasaki p.b.c.}, we depict a collection of the results of our
calculations of threshold values of $U$, obtained for perturbed flat-band sparse $\Delta$-chains. The rigorous upper bounds,
$\overline{U}_{th}$ versus $1/N_{{\cal{S}}}$ (see Tab.~\ref{sparse delta-chain, tasaki method} for numerical values),
obtained by Tasaki method, constitute a set of reference data.
This is a strictly increasing sequence, with a tendency to saturate for small values of $1/N_{{\cal{S}}}$;
very much like in the case of $\Delta$-chain.

The threshold values $U_{th}^{s.s.}$ for periodic boundary condition versus $1/N_{s}$
seem to oscillate around a constant value, irrespectively of whether the system is perturbed in Ichimura et al way or Tasaki way.
The numerical values of these data can be found in Tabs. ~\ref{small sparse delta-chain, ichimura et al pert., p.b.c.},
\ref{small sparse delta-chain, tasaki pert., p.b.c.}. In both cases, of Ichimura et al perturbation and Tasaki one, the data
are below the rigorous upper bounds.

\begin{figure}[H]
\begin{center}
   \includegraphics[scale=0.35]{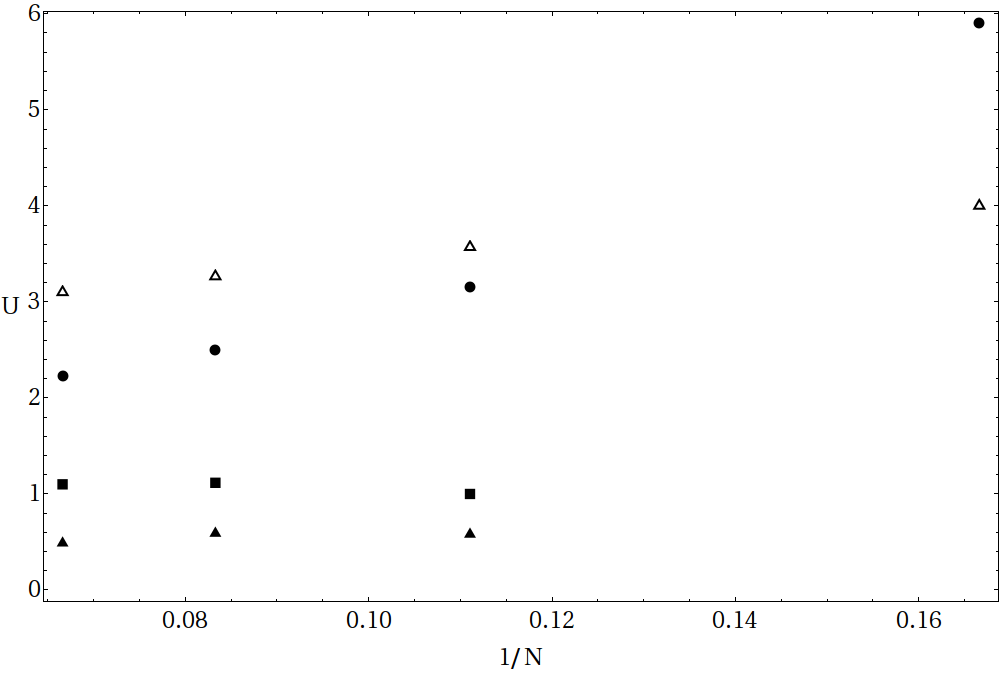}
%    \vspace{-80pt}
    \caption{\label{potential, ichimura p.b.c. and o.b.c., tasaki p.b.c.}
    The case of a nearly-flat-band sparse $\Delta$-chain. The hoping intensities of
    the unperturbed flat-band system are: $m=0.80$, $t=\sqrt{0.80}$.
    For the system perturbed in Ichimura et al way, with $m=0.80$ and $t=1.0$, $U_{th}^{s.s.}$ versus $1/N_s$:
    open triangles -- open-boundary condition with $N_s=6,9,12,15$ and $N_e=2,3,4,5$, respectively;
    solid triangles -- periodic-boundary condition with $N_s=9,12,15$ and $N_e=3,4,5$, respectively.
    For the system perturbed in Tasaki way, with periodic boundary condition, $m=0.80$, $t=\sqrt{0.80}$ and $s=0.07$
    (for such a value of $s$ the widths of the flat-band broadened by Tasaki and Ichimura et al perturbations
    are approximately the same), $U_{th}^{s.s.}$ versus $1/N_s$, with $N_s=9,12,15$ and $N_e=3,4,5$, respectively,
     -- solid squares.
    $\overline{U}_{th}$ versus $1/N_{{\cal{S}}}$, with $s=0.07$ and the underlying potentials defined in
    Figs.~\ref{one-W-valley potential},\ref{two-W-valley potential},
    obtained for $N_{{\cal{S}}}=6,9,12,15$ -- bullets. For the numerical values of depicted data see
    Tables.~\ref{small sparse delta-chain, ichimura et al pert., p.b.c.}, \ref{small sparse delta-chain, tasaki pert., p.b.c.},
\ref{sparse delta-chain, tasaki method}.}
\end{center}
\end{figure}

In contrast, the small-system threshold values of $U$ obtained for open-boundary condition are definitely to high. In the case
of Ichimura et al perturbation, the depicted in Fig.~\ref{potential, ichimura p.b.c. and o.b.c., tasaki p.b.c.} data
(which reproduce  the results of \cite{Ichimura-98}) seem to scale linearly with $1/N_s$.
However, except the case of 6 sites, their values are greater than the corresponding rigorous upper bounds provided
by Tasaki method. Moreover, also the value of a least-square linear extrapolation to $1/N_s \to 0$ limit, is above the rigorous
upper bound obtained from a potential whose support consists of 15 sites. The Ichimura et al choice of the hopping parameters
results in a nearly-flat-band width of the order $10^{-1}$ of the energy unit. We have repeated the calculations for another
values of hopping parameters that give the nearly-flat-band width of the order $10^{-2}$. As expected, the calculated values of
$U_{th}^{s.s.}$ are smaller, they also seem to scale linearly with $1/N_s$, but the slope of the least-square linear fit is smaller
than in the previous case, so that the value of a least-square linear extrapolation to $1/N_s \to 0$ limit is higher than in the
previous case.
In the case of Tasaki perturbation (open boundary condition) the values of $U_{th}^{s.s.}$ are even higher than in the case of
Ichimura et al perturbation
(see Tab.~\ref{small sparse delta-chain, tasaki pert., o.b.c.}).

Apparently, the linear scaling, observed in the case of the flat-band sparse $\Delta$-chain with Ichimura et al perturbation,
is accidental, and the predicted value of $U_{th}$ is not reliable.
Taking into account our results for the $\Delta$-chain, one can claim that small-system and open-boundary condition threshold
values of $U$ are irrelevant for our task of estimating $U_{th}$ of nearly-flat-band systems.
Definitely, the threshold values of $U$ obtained for periodic boundary conditions constitute  a better basis for deriving
reliable estimates of $U_{th}$.

\section{Summary}
Nearly-flat-band ferromagnets constitute a specific class of strongly-correlated electrons systems. Theoretically,
they can be thought of as weak perturbations of the underlying unphysical flat-band systems. One of their
characteristic features, which is a result of the competition between the kinetic energy and the
Coulomb interaction energy, is the threshold value $U_{th}$ of the screened Coulomb repulsion, above which a
nearly-flat-band electron system becomes ferromagnetic. This quantity is well defined for sufficiently large systems,
presumably not necessarily as large as macroscopic ones. It is desirable to have a reasonable estimate of that intrinsic
property of nearly-flat-band ferromagnets.

In this paper, we have made attempts at estimating $U_{th}$ in two nearly-flat-band systems, modeled by Hubbard
Hamiltonians (\ref{hubbard}), the nearly-flat-band $\Delta$-chain and the nearly-flat-band sparse $\Delta$-chain,
defined in Section 3 and Section 4, respectively.
This choice has been convenient, since it enabled us to avoid large volume computer calculations. Concerning the electric
conductivity both models can be classified as insulators or semiconductors.

We have considered two methods of estimating $U_{th}$, described in detail in Section 2. One of the methods,
the Tasaki method, provides us with rigorous upper bounds for true $U_{th}$, independent of the size of the system
and boundary conditions.
We have shown how to obtain the best bounds, whose quality is limited only by available computer facilities,
and obtained those bounds for both models. In the case of the sparse $\Delta$-chain, this resulted in a computer assisted
proof of nearly-flat-band ferromagnetism (the existence of flat-band ferromagnetism was demonstrated in \cite{Ichimura-98}).

The other method, studied by us, is the method of small-system estimates of $U_{th}$; this method amounts to
what a physicist would typically do when faced with such a problem. It provides us with boundary condition and perturbation
dependent threshold values $U_{th}^{s.s.}$. Then, the problem we have to deal with is how to extract reliable estimates
of  $U_{th}$, having a set of values of $U_{th}^{s.s.}$ obtained for systems of different sizes, open- or periodic-boundary
conditions, and various perturbations, like Tasaki or Ichimura et al perturbations.
Our main conclusion is that when applying the method of small-system estimates one should resort to periodic-boundary.
Concerning the perturbation, the Tasaki perturbation is preferable.

Finally, concerning the question of the range of values taken by $U_{th}$, let us note that, in the two models considered,
we have chosen one of the hopping intensities as the energy unit. The remaining independent hopping intensities have been
chosen to differ from this unit by not more than a few ten per cent. Then, the widths of the broadened flat-bands do not
exceed 14 percent of the energy unit. The resulting estimates of $U_{th}$ are either smaller or almost equal to the energy
unit. However, for the widths of the broadened flat bands that are about a few per cent of the energy unit, $U_{th}$ is
significantly smaller than the energy unit.

\section{Acknowledgements}
We are grateful to Andreas Stathopulos (Computer Science Department, College of William \& Mary, Williamsburg)
for guiding us through his C library {\em Primme} \cite{Primme},\cite{Diagonalization 1},
\cite{Diagonalization 2}.

\section{Appendix}

\begin{table}[H]
\begin{center}
%\begin{Large}
\begin{tabular}{|c|c|c|c|c|}\hline
$N_s$&6&8&10&12\\\hline
$N_e$&3&4&5&6\\\hline
$U_{th}$&0.104&0.109&0.105&0.110\\\hline
\end{tabular}
%\end{Large}
\end{center}
\caption{\label{small delta-chain, Tasaki pert.,p.b.c.}
$U_{th}^{s.s.}$ for $\Delta$-chains with periodic-boundary condition and Tasaki perturbation; $t_{+}=1.0$, $t_{-}=1.5$, $s=0.01$.
These data are depicted in Figs.\ref{potential, p.b.c., o.b.c.}, \ref{potential, p.b.c.} (solid squares).
The least-square estimate of a constant is 0.107.}
\end{table}

\begin{table}[H]
\begin{center}
%\begin{Large}
\begin{tabular}{|c|c|c|c|}\hline
$N_{s}$&9&12&15\\\hline
$N_e$&3&4&5 \\\hline
$U_{th}$&0.592&0.602&0.500\\\hline
\end{tabular}
%\end{Large}
\end{center}
\caption{\label{small sparse delta-chain, ichimura et al pert., p.b.c.}
$U_{th}^{s.s.}$ for nearly-flat-band sparse $\Delta$-chains with periodic-boundary condition and
Ichimura et al perturbation; $m=0.8$, $t=1.0$.
These data are depicted in Fig.\ref{potential, ichimura p.b.c. and o.b.c., tasaki p.b.c.} (solid triangles).
The least-square estimate of a constant is 0.565. }
\end{table}

\begin{table}[H]
\begin{center}
%\begin{Large}
\begin{tabular}{|c|c|c|c|}\hline
$N_s$&9&12&15\\\hline
$N_e$&3&4&5\\\hline
$U_{th}$&0.994&1.110&1.094\\\hline
\end{tabular}
%\end{Large}
\end{center}
\caption{\label{small sparse delta-chain, tasaki pert., p.b.c.}
$U_{th}^{s.s.}$ for nearly-flat-band sparse $\Delta$-chains with periodic-boundary condition and
Tasaki perturbation; $m=0.8$, $t=\sqrt{0.8}$, $s=0.07$.
These data are depicted in Fig.\ref{potential, ichimura p.b.c. and o.b.c., tasaki p.b.c.} (solid squares).
The least-square estimate of a constant is 1.066.}
\end{table}

\begin{table}[H]
\begin{center}
%\begin{Large}
\begin{tabular}{|c|c|c|c|c|}\hline
$N_s$&6&9&12&15\\\hline
$N_e$&2&3&4&5\\\hline
$U_{th}$&5.900&3.147&2.498&2.225\\\hline
\end{tabular}
%\end{Large}
\end{center}
\caption{\label{sparse delta-chain, tasaki method}
$\overline{U}_{th}$ obtained from nearly-flat-band sparse $\Delta$-chains; $m=0.8$, $t=\sqrt{0.8}$, $s=0.07$.
These data are depicted in Fig.\ref{potential, ichimura p.b.c. and o.b.c., tasaki p.b.c.} (bullets).}
\end{table}

\begin{table}[H]
\begin{center}
%\begin{Large}
\begin{tabular}{|c|c|c|c|c|c|c|c|c|}\hline
$N_s$&6&6&9&9&12&12&15&15\\\hline $N_e$&2&3&3&4&4&5&5&6\\\hline
$U_{th}$&22.030&4.613&8.986&7.71&8.838&10.195&9.290&11.016\\\hline
\end{tabular}
%\end{Large}
\end{center}
\caption{\label{small sparse delta-chain, tasaki pert., o.b.c.}
$U_{th}^{s.s.}$ for small nearly-flat-band sparse $\Delta$-chains with open-boundary condition and
Tasaki perturbation; $m=0.8$, $t=\sqrt{0.8}$, $s=0.07$.}
\end{table}

%\newpage

\end{document}